%% file: main.tex
\documentclass{article}
\usepackage{template/dcase2024}
\usepackage{amsmath, amssymb,times,booktabs, tabularx}
\usepackage[colorlinks=true,allcolors=black]{hyperref}
\usepackage[nameinlink,capitalize]{cleveref} %
\crefname{figure}{Figure}{Figure} %
\Crefname{equation}{Equation}{Equations} %
\renewcommand\autoref{\cref}

\usepackage{csquotes}
\usepackage{multirow, multicol, makecell}
\usepackage{mathtools}
\DeclarePairedDelimiter\round{\lfloor}{\rceil}

\DeclareMathSymbol{\uminus}{\mathbin}{AMSa}{"39}
\newcommand{\degree}{^ {\circ}}
\newcommand{\DOAE} {DO\hspace{-.1em}AE} %
\newcommand{\DOA}{DO\hspace{-.1em}A}

\title{Learning Multi-Target TDOA Features \\ for Sound Event Localization and Detection}

\input{authors}

\begin{document}

\ninept
\maketitle

\begin{abstract}
Sound event localization and detection (SELD) systems using audio recordings from a microphone array rely on spatial cues for determining the location of sound events. As a consequence, the localization performance of such systems is to a large extent determined by the quality of the audio features that are used as inputs to the system. We propose a new feature, based on neural generalized cross-correlations with phase-transform (NGCC-PHAT), that learns audio representations suitable for localization. Using permutation invariant training for the time-difference of arrival (TDOA) estimation problem enables NGCC-PHAT to learn TDOA features for multiple overlapping sound events. These features can be used as a drop-in replacement for GCC-PHAT inputs to a SELD-network. We test our method on the STARSS23 dataset and demonstrate improved localization performance compared to using standard GCC-PHAT or SALSA-Lite input features. 
\end{abstract}


\section{Introduction}
\label{sec:intro}

\begin{figure}[t!]
    \centering
    \includegraphics[trim={2cm 0 2cm 0},clip,width=\linewidth]{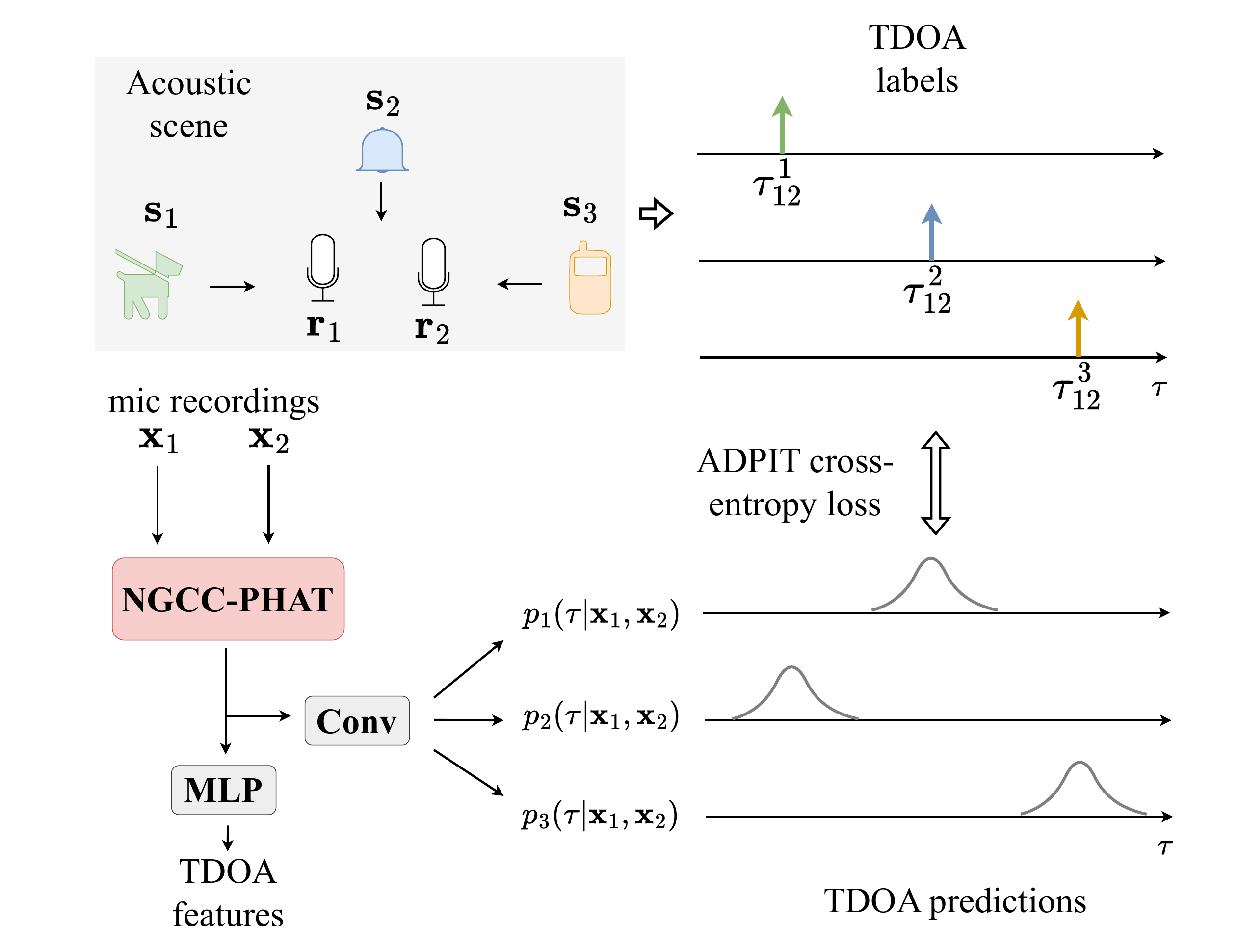}
    \vspace{-0.5cm}\caption{Overview of our pre-training strategy with $K=3$ tracks. Given a set of sound events, we train a neural GCC-PHAT to predict the TDOA of each event. When the number of sound events is less than $K$, auxiliary duplication of the labels is used. In this illustration, only two microphones are shown for brevity.}
    \label{method-overview}
\end{figure}

The sound event localization and detection (SELD) task consists of classifying different types of acoustic events, while simultaneously localizing them in 3D space. The DCASE SELD Challenge \cite{DCASE2024_seld_challenge} provides first order ambisonics (FOA) recordings and signals captured from a microphone array (MIC). In recent years, most systems submitted to the challenge have utilized the former format,  whereas the latter has been less explored. In this work, we therefore focus on how to better exploit information in the MIC recordings by learning to extract better features.

Generalized cross-correlations with phase transform (GCC-PHAT) \cite{knapp1976generalized} combined with spectral audio features is the basis for most SELD methods for microphone arrays. The spectral features contain important cues on what type of sound event is active, whereas the purpose of GCC-PHAT is to extract the time-differences of arrival (TDOA) for pairs of microphones. The TDOA measurements can then be mapped to direction-of-arrival (DOA) estimates, given the geometry of the array. However, GCC-PHAT is known to be sensitive to noise and reverberation \cite{champagne1996performance}. GCC-PHAT can also fail to separate TDOAs for overlapping events, since two events at different locations can have the same TDOA for a given microphone pair, which yields only one correlation peak.

To improve separation of overlapping events, Xu et al.\ \cite{xue20b_interspeech} proposed a beamforming approach, where phase differences from the cross-power spectrum are used as input features. Similarly, Cheng et al.\ \cite{cheng2023improving} showed that localization performance can be improved by first filtering the audio signals using a sound source separation network before performing feature extraction. Several works \cite{sundar, he22_interspeech} have also proposed end-to-end localization from raw audio signals. The most widely adopted input feature is however the spatial cue-augmented log-spectrogram (SALSA) \cite{nguyen2022salsa} and variants thereof (SALSA-Lite) \cite{nguyen2022salsalite}, that combine directional cues with spectral cues in a single feature. This is done by calculating the principal eigenvector of the spatial covariance matrix for the different frequencies in the spectrogram. 

Although some recent works \cite{xiao2015learning, salvati2021time, berg22_interspeech} have approached TDOA estimation using learning-based methods, there is a lack of research in how to combine this with the SELD task. Berg et al.\ \cite{berg22_interspeech} proposed using a shift-equivariant neural GCC-PHAT (NGCC-PHAT) network. However, this method, as it was originally proposed, only supports single-source TDOA estimation and was not evaluated in a real-world localization scenario. 

In this work, we describe how NGCC-PHAT can be trained to extract TDOA features for multiple sound sources. We show that such features can be learnt by employing permutation invariant training, which allows for prediction of TDOAs for multiple overlapping sound events. Furthermore, we show that these features can be used with an existing SELD-pipeline on a real-world dataset, for better performance compared to using traditional input features. The material presented in this work is an extension of our DCASE 2024 challenge submission \cite{Berg_LU_task3_report}.

\section{Method}
\label{sec:method}
\subsection{Background}
\label{sec:method1}

Consider an acoustic scene, as shown in Figure \ref{method-overview}, with $M$ microphones located at positions $\mathbf{r}_m \in \mathbb{R}^3$ for $m = 1, \hdots, M$. Furthermore, let $\mathbf{s}_p \in \mathbb{R}^3$, $p = 1, \hdots, P$ denote the locations of the active sound events. For a given time frame, each microphone records a signal $x_i$, which is composed of the sum of active events as
\begin{align}
    x_i[n] = \sum_{p=1}^P (h_{p,i} * u_p) [n] + w_i[n] , \quad n=1,\hdots, N,
\end{align}
where $u_p$ is the $p$:th active event, $h_{p,i}$ is the room impulse response from the $p$:th event to the $i$:th microphone, $w_i$ is additive noise and $N$ is the number of samples. Furthermore, we define the TDOA for microphone pair $(i,j)$ and the $p$:th event as 
\begin{align}
    \tau^p_{ij} = \round{ \frac{F_s}{c} \left( ||\mathbf{s}_p - \mathbf{r}_i||_2 - ||\mathbf{s}_p - \mathbf{r}_j||_2  \right) },
\end{align}
where $F_s$ is the sampling rate, $c$ is the speed of sound and $\round{\cdot}$ denotes rounding to the nearest integer.

The GCC-PHAT is defined as
\begin{align}
\label{gcc-phat}
R_{ij}[\tau] = \frac{1}{N} \sum_{k=0}^{N-1}  \frac{X_i[k]X_j^*[k]}{|X_i[k]X_j^*[k]|}e^{\frac{i2\pi k\tau}{N}},
\end{align}
where $(X_i, X_j)$ are the discrete Fourier transforms of $(x_i, x_j)$. The feature is calculated for time delays $\tau = -\tau_{\text{max}}, ..., \tau_{\text{max}}$, where $\tau_{\text{max}} = \max_{i,j} \round{||\mathbf{r}_i-\mathbf{r}_j||_2 F_s / c}$ is the largest possible TDOA for any pair of microphones. In an anechoic and noise-free environment with a single sound event $u_p$, this results in $R_{ij}[\tau] = \delta_{\tau_{ij}^p}[\tau]$, where
\begin{align}
    \delta_{\tau_{ij}^p} [\tau] = \begin{cases}
        1, \quad \tau = \tau_{ij}^p, \\
        0, \quad \text{otherwise.}
    \end{cases}
\end{align}

In practice, GCC-PHAT will often yield incorrect TDOA estimates due to noise and reverberation. In the case of multiple overlapping sound events, the different events may interfere and result in difficulties resolving peaks in their signal correlations.

NGCC-PHAT attempts to alleviate this problem by filtering the input signals using a learnable filter bank with $L$ convolutional filters, before computing GCC-PHAT features $R_{ij}^l$, $l=1,\hdots,L$ for each channel in the signals independently. In theory, such a filter bank can perform source separation so that different channels in the NGCC-PHAT correspond to TDOAs for different sound events. Note that for an ideal filter bank that perfectly separates the $p$:th sound event to the $l$:th channel, we would have $R_{ij}^l[\tau] =\nolinebreak\delta_{\tau_{ij}^p}[\tau]$ in an anechoic and noise-free environment, due to the shift-equivariance of the convolutional filters.

\subsection{Permutation Invariant Training for TDOA Estimation}
\label{sec:method2}

We extend NGCC-PHAT to predict time delays for multiple events in a single time frame using auxiliary duplicating permutation invariant training (ADPIT) \cite{shimada2022multi}, by creating separate target labels for each active sound event. This is done by training a classifier network to predict the TDOA of all active events for all pairs of microphones by treating it as a multinomial classification problem. The $L$ correlation features are first processed using another series of convolutional layers with $C$ output channels. These are then projected to $K$ different output tracks which are assigned to the different events. The last layer of the NGCC-PHAT network therefore outputs probability distributions $p_k(\tau | \mathbf{x}_i, \mathbf{x}_j )$ for $k=1,\hdots,K$ over the set of integer delays $\tau \in \{-\tau_{\text{max}}, ..., \tau_{\text{max}} \}$, as illustrated in Figure \ref{method-overview}.

With $K$ as the number of tracks, assume for now that there are also $P = K$ active events. Furthermore, let $\text{Perm}([K])$ denote the set of permutations of the events $\{1, \hdots, K\}$. For a single microphone pair $(i,j)$ and an event arrangement $\alpha \in \text{Perm}([K])$, the loss is calculated using the average cross-entropy over all output tracks as 
\begin{align}
    l_\alpha(\mathbf{x}_i, \mathbf{x}_j) = - \frac{1}{K} \sum_{k=1}^K \sum_{\tau=\uminus\tau_{\text{max}}}^{\tau_{\text{max}}} \delta_{\tau_{ij}^{\alpha (k)}} [\tau] \log p_k(\tau | \mathbf{x}_i, \mathbf{x}_j ).
\end{align}
Due to the ambiguity in assigning different output tracks to different events, we calculate the loss for all possible permutations of the events and use the minimum. The loss is then averaged over all $M(M-1)/2$ microphone pairs, giving the total loss 
\begin{align}
    \mathcal{L} = \frac{2}{M(M-1)}\sum_{{i,j=1}\atop{i < j}}^M \min_{\alpha \in \text{Perm}([K])} l_\alpha(\mathbf{x}_i, \mathbf{x}_j). 
\end{align}
Note that this loss function is class-agnostic, since the output tracks are not assigned class-wise. The main purpose of the TDOA features are therefore to provide better features for localization when combined with spectral features that are suitable for classification. 

When the assumption $P = K$ does not hold, the formal implication is that $\alpha$ needs to cover another set of event arrangements. Our approach is equivalently to transform each such case into subcases where the assumption holds. Time frames with no active events ($P=0$) are discarded in the loss calculation, since no TDOA label can be assigned. When $1 \leq P \leq K-1$, we perform auxiliary duplication of events following the method in \cite{shimada2022multi}, which makes the loss invariant to both permutations and which events that are duplicated. Furthermore, in the case of $K < P$, it is possible to compute the loss for all subsets of $K$ events from $P$ and use the minimum.

\section{Experimental Setup}
\label{sec:setup}

\subsection{Using TDOA Features for SELD}
\label{sec:setup1}

In order to show the benefits of better TDOA features for SELD, we demonstrate how they can be used in conjunction with a SELD-system. This involves two training phases: 1) pre-training of the NGCC-PHAT network for TDOA prediction and 2) training the SELD-network using the TDOA features as input. The NGCC-PHAT network operates on raw audio signals and consists of four convolutional layers, the first being a SincNet \cite{ravanelli2018speaker} layer, and the remaining three use filters of length 11, 9, and 7 respectively. Here, each convolutional layer has $L=32$ channels and together form the filter bank mentioned in \autoref{sec:method1}, which is applied independently to audio from the different microphones. GCC-PHAT features are computed channel-wise for each microphone pair, and the features are then processed by another four convolutional layers, where the final layer has $C=16$ output channels. 

The maximum delay used is chosen for compatibility with the setup in the STARSS23 dataset \cite{Shimada2023starss23}, which uses a tetrahedral array with $M=4$ microphones. The diameter of the array is 8.4~cm, which corresponds to a maximum TDOA of $\tau_{\text{max}} = 6$ delays at a sampling rate of $F_s = 24$ kHz. In total, the TDOA features therefore have shape $[C, M(M-1)/2,  2\hspace{.1em}\tau_{\text{max}}+1] = [16, 6, 13]$.

During pre-training for TDOA-prediction, the 16 channels are then mapped by a convolutional layer to $K=3$ output tracks. Although the maximum polyphony in a single time frame in the dataset is five, we use $K=3$ tracks since the computational complexity of permutation invariant training scales as $\mathcal{O}(K!)$ and more than three simultaneous events are rare. When more than three events are active, for pre-training we randomly select labels for three events and discard the rest. 

When training the SELD-network, we extract the TDOA input features for longer audio signals by windowing the NGCC-PHAT computation without overlap. We use an input duration of 5 second audio inputs, which corresponds to $T = 250$ TDOA features when using a window length of 20 ms. Since the TDOA features are designed to be class-agnostic, we combine them with spectral features for the same time-frame in order to better distinguish between different types of event. For this we use log mel-spectograms (MS) with $F = 64$ spectral features for each recording.

\begin{figure}[t!]
    \centering
    \includegraphics[trim={4cm 0.2cm 1cm 0},clip, width=\linewidth]{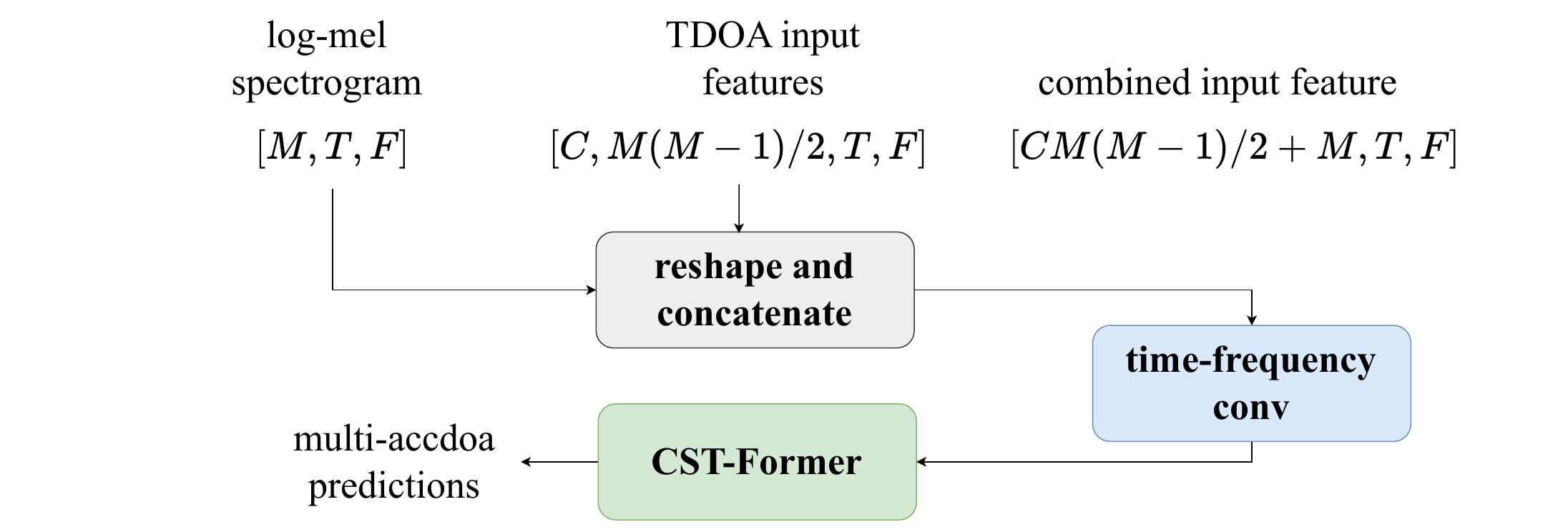}
    \caption{Illustration of how TDOA features are used together with log mel-spectrograms as input to the CST-Former network.}
    \label{method-flow}
\end{figure}

When merging the spectral features with the TDOA features, we first concatenate the 16 channels for the 6 microphone pairs of the TDOA features, and use a multi-layer perceptron to map the 13 time-delays to 64 dimensions. The TDOA features are then reshaped and concatenated with the $M$ spectral features channel-wise, as shown in Figure \ref{method-flow}, resulting in a combined feature size of $[CM(M-1)/2+M, T, F] = [100, 250, 64]$. 

The combined feature is passed through a small convolutional network with 64 output channels with pooling over the time and spectral dimensions. Here we use two pooling variants that determine the size of the input features to the SELD-network: 1) pooling over 5 time windows and 4 frequencies, which produces features of size [64, 50, 16], or 2) pooling over 5 time windows and no pooling over frequencies, which results in features of size [64, 50, 64]. We call the resulting network variants \textit{Small} and \textit{Large} for this reason.

For SELD-training, we use a CST-Former \cite{shul2024cst} network that consists of Transformer blocks, where each block contains three self-attention modules: temporal attention, spectral attention and channel attention with unfolded local embedding. We use the default configuration with two blocks, each with eight attention heads, and refer to \cite{shul2024cst} for more details about this architecture.
\subsection{Dataset and Model Training}
We train all our models on a mixture of real spatial audio recordings and simulated recordings. The real recordings are from the STARSS23 \cite{Shimada2023starss23} audio-only dev-train dataset, which consists of about 3.5 hours of multi-channel audio recordings. The dataset has up to 5 simultaneous events from 13 different classes. For data augmentation, we use channel-swapping \cite{wang2023four}, which expands the dataset by a factor of 8 by swapping the input channels and corresponding DOA labels in different combinations. 

The simulated data is provided as a part of the DCASE 2024 challenge \cite{krause_2024_10932241} and consists of 20 hours of synthesized recordings, where the audio is taken from the FSD50K \cite{fonseca2021fsd50k} dataset. In addition, we generate an additional 2 hours of synthesized recordings using Spatial Scaper \cite{roman2024spatial} with impulse responses from the TAU \cite{politis_2022_6408611} and METU \cite{orhun_olgun_2019_2635758} databases. This additional data contains sounds from classes that occur rarely in STARSS23, namely \enquote{bell}, \enquote{clapping}, \enquote{doorCupboard}, \enquote{footsteps}, \enquote{knock} and \enquote{telephone}. The total amount of training data is about 50 hours. 

The NGCC-PHAT network was trained for one epoch with a constant learning rate of 0.001, after which the weights were frozen. The CST-Former network was then trained for 300 epochs using the AdamW optimizer \cite{loshchilov2017decoupled} with a batch size of 64, a cosine learning rate schedule starting at 0.001 and weight decay of 0.05. The mean squared error was used as loss function with labels in the Multi-ACCDOA \cite{shimada2022multi} format, with distances included as proposed in \cite{krause2024sound}. In order to penalize errors in predicted distance relative to the proximity of the sound events, we scale loss-terms for the distance error with the reciprocal of the ground truth distance.

\begin{figure*}[t]
    \centering
    \includegraphics[trim={0cm 0.5cm 0cm 0.1cm},clip,width=\linewidth]{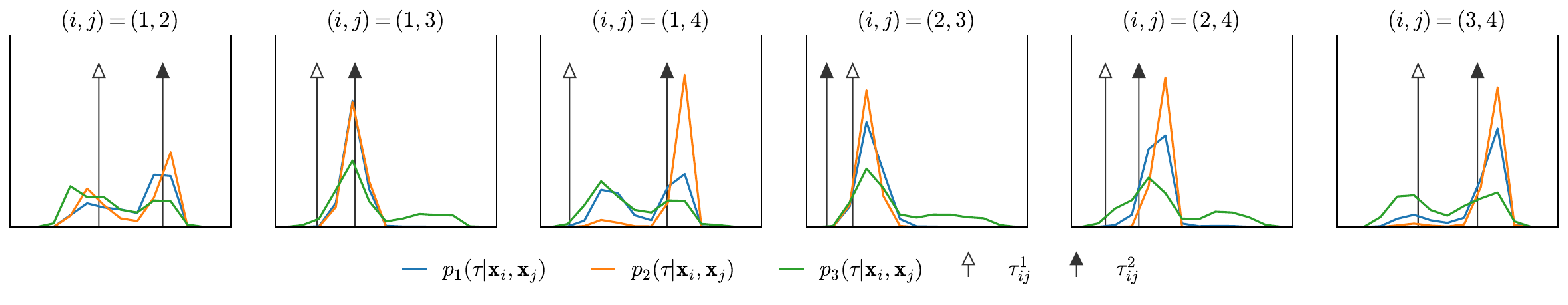}
    \vspace{-0.5cm}\caption{An example of the TDOA predictions $p_k(\tau | \mathbf{x}_i, \mathbf{x}_j )$ from the pre-trained NGCC-PHAT network using $K=3$ output tracks. Predictions are shown for all six microphone combinations $(i,j)$ at a single time frame with two events and ground truth TDOAs $\tau_{ij}^1$ and $\tau_{ij}^2$.}
    \label{tdoa-fig}
\end{figure*}

Evaluations were done using the DCASE 2024 SELD challenge metrics \cite{DCASE2024_seld_challenge, politis2020overview}. This includes the location dependent F-score $F_{LD}$, the DOA error $\DOAE$ and the relative distance error $RDE$, which is the distance error divided by the ground truth distance to the event. Each metric is calculated class-wise and then macro-averaged across all classes. Furthermore, the location dependent F-score only counts predicted events as true positives if they are correctly classified and localized, such that predictions with $\DOAE$ larger than $T_{\DOA} = 20 \degree$ or $RDE$ larger than $T_{RD} = 1$ are counted as false positives. We focus on evaluating the performance of our method compared to that of other commonly used input features with the same SELD-network, and do not compare to other (e.g.\ FOA-based) state-of-the-art methods.

\section{Results}

Our main results are presented in Table \ref{results}, where we compare our method to GCC with MS and to SALSA-Lite. Our method performs better in terms of $F_{LD}$ and $\DOAE$, for both the Small and Large variant of the network, although SALSA-Lite has the lowest $RDE$ for the Large variant. When increasing the model size, the results improve for both SALSA-Lite and NGCC, but not for GCC. Since GCC features are less informative, the increase in model size results in overfitting. The same can be said for the increase in $RDE$ when using NGCC + MS, since the TDOA features from both GCC and NGCC mostly contain angular cues, but less information about spatial distance. Note that GCC + MS and NGCC + MS use exactly the same CST-Former architecture, so the extra parameter count when using NGCC comes from the pre-trained feature extractor. When using SALSA-Lite, the pooling operations in the convolutional layers were adjusted in order to achieve a similar model size.
\begin{table}[t!]
    \centering
    \vspace{-0.5cm}\caption{Macro-averaged test results on STARSS23 \cite{Shimada2023starss23} dev-test.}
    \label{results}
    \resizebox{\linewidth}{!}{
    \begin{tabular}{l|cccc} \toprule 
    Input feature & $F_{LD} \uparrow$ & $DO\hspace{-.1em}AE \downarrow$ & $RDE \downarrow$ & \#params \\  \midrule
    \multicolumn{5}{c}{CST-Former Small} \\ \midrule
    GCC + MS & $15.7 \pm 1.0$ & $27.7 \pm 2.1$ & $0.78 \pm 0.02$ & 550K \\
    SALSA-Lite & $24.6 \pm 2.0$ & $27.0 \pm 1.2$ & $\mathbf{0.41 \pm 0.02}$ & 530K \\
    NGCC + MS & $\mathbf{26.0 \pm 2.0}$ & $\mathbf{25.8 \pm 2.3}$ & ${0.42 \pm 0.01}$ & 663K \\ \midrule
    \multicolumn{5}{c}{CST-Former Large} \\ \midrule
    GCC + MS & $14.2 \pm 1.1$ & $28.4 \pm 1.9$ & $0.84 \pm 0.03$ & 1.37M \\
    SALSA-Lite & $26.1 \pm 1.0$ & $26.4 \pm 3.6$ & $\mathbf{0.42 \pm 0.02}$& 1.35M \\
    NGCC + MS & $\mathbf{28.2 \pm 2.8}$ & $\mathbf{23.2 \pm 1.8}$ & $0.50 \pm 0.02$ & 1.49M \\
    \bottomrule
    \end{tabular}
    }
\end{table}
\begin{table}[t]
    \centering
    \vspace{-0.2cm}\caption{Ablations of the number input channels used in the TDOA input features for CST-Former Small.}
    \label{channel-ablation}
    \footnotesize
    \begin{tabular}{l|cccc} \toprule 
    $C$ & $F_{LD} \uparrow$ & $DO\hspace{-.1em}AE \downarrow$ & $RDE \downarrow$ & \#params \\  \midrule
    1 & $24.4 \pm 2.3$ & $29.7 \pm 3.3$ & $0.44 \pm 0.08$ & 608K  \\
    4 & $24.2 \pm 0.8$ & $\mathbf{23.2 \pm 2.5}$ & $0.46 \pm 0.01$ & 619K\\
    16 & $\mathbf{26.0 \pm 2.0}$ & $25.8 \pm 2.3$ & $\mathbf{0.42 \pm 0.01}$ & 663K \\ \bottomrule
    
    \end{tabular}
\end{table}

\begin{figure}[t]
    \centering
    \vspace{-0.2cm}\includegraphics[trim={0cm 0.3cm 0cm 0.3cm},clip,width=\linewidth]{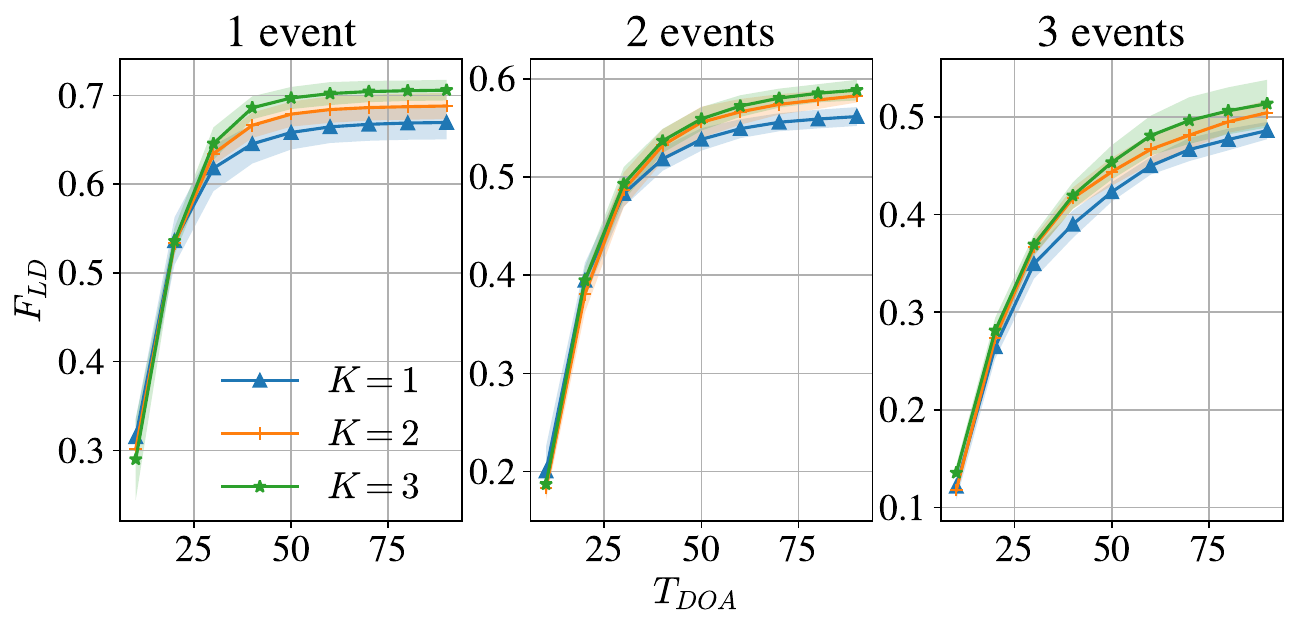}
    \vspace{-0.5cm}\caption{Micro-averaged F-score as a function of the angular threshold $T_{\DOA}$ using different number of output tracks $K$ during TDOA pre-training. Evaluation was done using CST-Former Small.}
    \label{track-ablation}
\end{figure}

In order to verify the importance of using more than one input channel for TDOA features, we ablate the number of channels $C$ in the NGCC-PHAT network. The results are shown in \autoref{channel-ablation}, where it can be seen that increasing the number of channels from 1 to 16 increases performance in terms of all metrics. This agrees with the intuition that using more than one input channels enables the pre-training to better separate spatial cues from different events. Furthermore, the cost for increasing the number channels in terms of the increase in model parameters is relatively small.

We also ablate the number of tracks $K$ used for TDOA-prediction during pre-training, and present the location dependent F-score for values of $T_{\DOA}$ in \autoref{track-ablation}. Due to the sensitivity of the macro-averaged F-score to incorrect predictions for rare classes in the test data, we instead use the micro-averaged statistic. At the default $20\degree$ threshold, the effect of increasing the number of tracks is small, but asymptotically it is clear that using $K=3$ tracks increases the F-score regardless of how many events are active. Note that the number of tracks only affects the complexity in the pre-training stage of NGCC-PHAT, and not the overall parameter count of the final model, since all $C$ channels are used as input to the network, and the mapping to $K$ tracks can be discarded.

Finally, we show examples of TDOA predictions in Figure \ref{tdoa-fig}. When the TDOAs of the events are well-separated, the different tracks yield different peaks at approximately the correct time delays. However, for the microphone pairs where events are tightly spaced, the predictions fail to separate the different TDOAs.

\section{Conclusions}
\label{sec:conclusions}

In this work we proposed an input feature based on NGCC-PHAT and showed its usefulness as input to a SELD-network. Permutation invariant training for the TDOA estimation
problem enabled NGCC-PHAT to learn TDOA features for multiple
overlapping sound events, and improved
SELD performance compared to using GCC-PHAT
or SALSA-Lite input features.

These results indicate that our NGCC-PHAT pre-training for TDOA classification provides a good feature extractor for the SELD task. Intuitively, better TDOA prediction in the feature extractor ought to yield better SELD results, but further studies are needed to validate this. Evaluating TDOA prediction performance would however involve new methodology, such as heuristics for peak selection from the output tracks, as well as selecting useful evaluation metrics. The downstream network could be resilient to some type of information our current loss function aims to suppress. In addition, a source-wise or class-wise TDOA format could be beneficial. We therefore anticipate future work to explore other pre-training options and end-to-end training.

Focusing on the feature extractor, we made minimal effort to address the other challenges of the dataset. 
We leave for future work to incorporate known techniques, such as class balancing, additional data augmentation, temporal filtering and ensemble voting.

\clearpage
\newpage
\begin{sloppy}

\bibliographystyle{template/IEEEtran}
\bibliography{dcase_refs}

\end{sloppy}
\end{document}

%% file: authors.tex
\name{Axel Berg$^{1,2}$\thanks{This work was partially supported by the strategic research project \mbox{ELLIIT} and the Wallenberg AI, Autonomous Systems and Software Program (WASP), funded by the Knut and Alice Wallenberg (KAW) Foundation. Model training was enabled by the Berzelius resource provided by the KAW Foundation at the National Supercomputer Centre in Sweden. \newline \newline
Code: \protect\url{https://github.com/axeber01/ngcc-seld/}},
      Johanna Engman$^{1}$,
      Jens Gulin$^{1,3}$, 
      Karl Åström$^{1}$, 
      Magnus Oskarsson$^{1}$,
      }
\address{$^1$Computer Vision and Machine Learning, Centre for Mathematical Sciences, \\Lund University, Sweden          
        \{firstname.lastname@math.lth.se\} \\
        $^2$ Arm, Lund, Sweden \{axel.berg@arm.com\}  \\ $^3$ Sony Europe B.V., Lund, Sweden \{jens.gulin@sony.com\}\\
 }